\documentclass[12pt]{article}
\usepackage{lineno}
\usepackage[margin=1in]{geometry}
\usepackage{amssymb,amsmath,latexsym}                        
\usepackage{graphicx}
\usepackage{cite}
\usepackage{xcolor}
\usepackage{float}

\usepackage{amsthm}
\usepackage{multicol}
\usepackage{authblk}
\usepackage[colorlinks]{hyperref}

\usepackage{adjustbox}
\usepackage{bbm}
\usepackage{mathtools}

\date{\today} 
\usepackage{graphicx}
\usepackage{subcaption}
\begin{document}

\title{\boldmath A neural network based algorithm for MRPC time reconstruction}
\author[a]{Fuyue Wang,}
\author[a]{Dong Han,}
\author[a]{Yi Wang,}
\author[a]{Yancheng Yu,}
\author[b]{Baohong Guo,}
\author[a]{Yuanjing Li}
\affil[a]{\normalsize\it Tsinghua University, \\Key Laboratory of Particle and Radiation Imaging, Ministry of Education, Beijing 100084, China}
\affil[b]{\normalsize\it Nuctech Company Limited}

\maketitle

\begin{abstract}
Multi-gap Resistive Plate Chamber(MRPC) is a widely used timing detector with a typical time resolution of about 60 $ps$. This makes MRPC an optimal choice for the time of flight(ToF) system in many large physics experiments. The prior work on improving the time resolution is mainly focused on altering the detector geometry, and therefore the improvement of the data analysis algorithm has not been fully explored. This paper proposes a new time reconstruction algorithm based on the deep neural networks(NN) and improves the MRPC time resolution by about 10 $ps$. Since the development of the high energy physics experiments has pushed the timing requirements for the MRPC to a higher level, this algorithm could become a potential substitution of the time over threshold(ToT) method to achieve a time resolution below 30 $ps$.  
\end{abstract}
\section{Introduction}
\label{sec:intro}
\indent Multi-gap resistive plate chamber(MRPC) is a well-preformed gasous detector that has been widely used in many large high energy physics experiments, such as Solenoidal Tracker at RHIC(STAR) and Compressed Baryonic Matter(CBM). In both of these two experiments, MRPC is an important element of the Time-of-flight(TOF) system and plays a crucial role in identifying different kind of the particles. The typical time resolution for present MRPC detectors is about 60 $ps$.

Prior work on improving the time resolution of MRPC mainly focuses on updating the material of the resistive plates or the gas and adjusting the detector geometry, while little work has been done on improving the time reconstruction algorithm. For present MRPCs, the signal is read out with the Time-Over-Threshold(ToT) method, which has a straightforward implementation in electronics. By setting a fixed threshold, this method measures the threshold crossing time $t_c$  and the total time over threshold $t_{tot}$\cite{nino:2004}. $t_c$ is regarded as the particle arriving time, and $t_{tot}$ which is strongly related to the pulse height is used to correct the time walk of $t_c$. An evident drawback of this method is that only two variables are extracted from the entire waveform, so the information is not enough for a higher demand of the resolution. $t_c$ and $t_{tot}$ are given in TDC channels, and the uncertainty of each channel is around 20$\sim$30 $ps$, which is large enough to overwhelm the intrinsic MRPC resolution. Besides, the correction of the time walk is complicated in large physics experiments which makes it hard to perform online. In this paper, we propose an end-to-end solution to obtain the MRPC time with an artificial neural network(NN). This NN algorithm takes the advantage of the entire leading edge of the signal and improves the time resolution by 10 $ps$ for the MRPC studied in this paper.

We simulate a thin gap (0.104mm thick) MRPC and study the time resolution with both methods mentioned above. The network is trained and validated with the simulation data, while test on both the simulation and experiment data. The simulation is described in Sec.\ref{sec:simu}. The structure of the neural network and the analysis algorithm are illustrated in Sec.\ref{sec:NN}. The results are shown in Sec.\ref{sec:resu}. 

\section{MRPC Simulation}
\label{sec:simu}
\begin{figure*}[h!]
	\centering
	\includegraphics[width=0.5\textwidth]{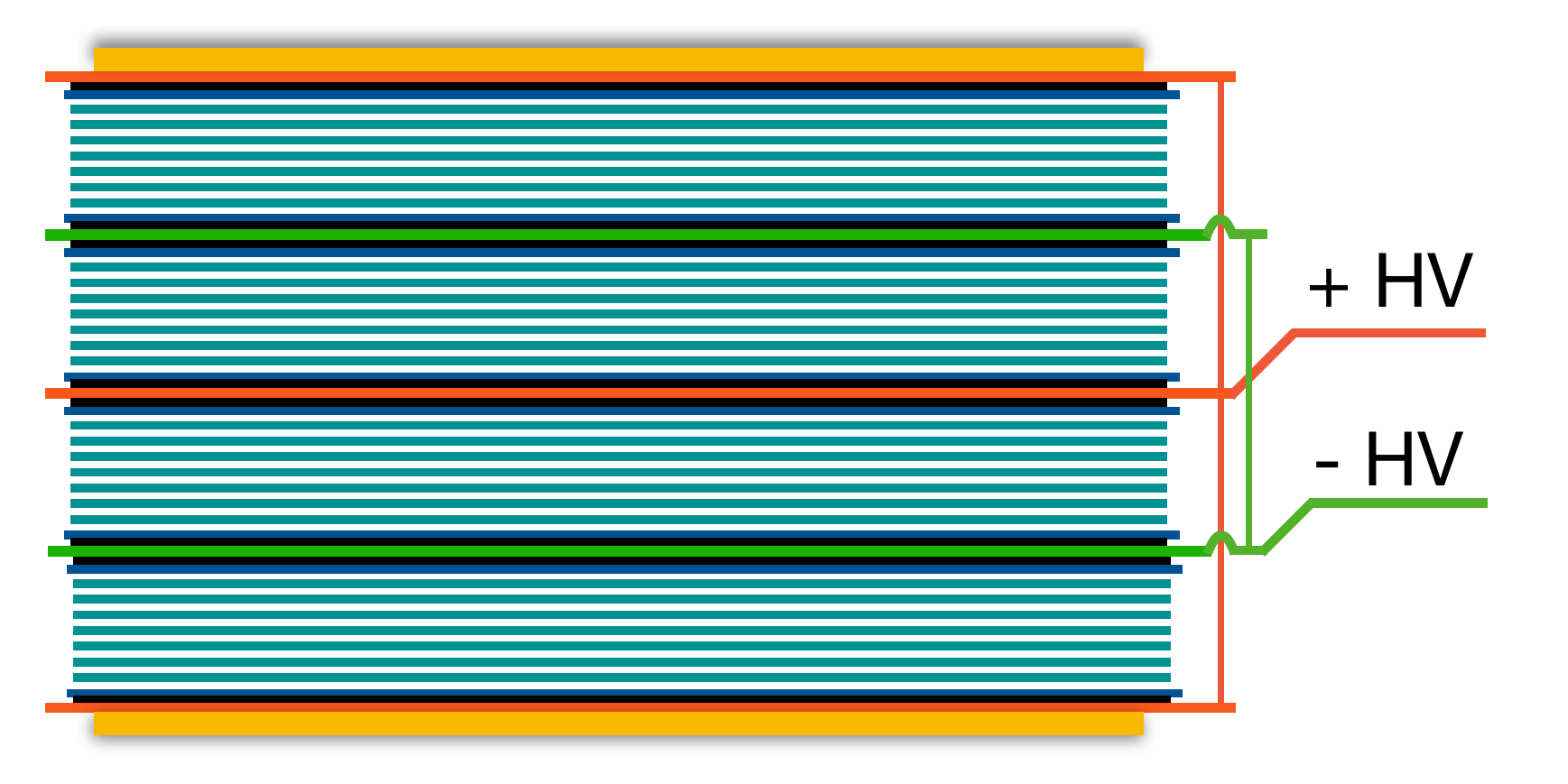}
	\caption{The structure of MRPC. It has 4 stacks, each with 8 gas gaps of 104 $\mu$m. }
	\label{fig:MRPCStru}
\end{figure*}

The simulation of MRPC is based on a standalone framework\cite{SimulationFuyue} built by our group. The structure of the MRPC studied in this paper is shown in Fig.\ref{fig:MRPCStru}. It has 4 stacks and 8 gas gaps in each stack. The thickness of the gaps and the resistive plates are 0.104 $mm$ and 0.5 $mm$ respectively. The working gas is the standard mixture 90\% $C_2H_2F_4$, 5\% $C_4H_{10}$ and 5\% $SF_6$ at room temperature and under standard atmosphere. The 1st, 3rd and 5th plates are connected to the positive high voltage, while the 2nd and 4th are negative.

Cosmic ray muons with a mean energy of 4 $GeV$ are simulated and used as the particle source. The angle of muons is largely perpendicular to the MRPC surface but with a small uncertainty. A fixed threshold is set to the simulated signal waveform and $t_c$, $t_{tot}$ are obtained. Another uncertainty of 25 $ps$ is added to these 2 variables which simulates the uncertainty of the TDC channel. 
\section{The neural network structure and time reconstruction algorithm}
\label{sec:NN}
Artificial neural network(NN) is a powerful tool for solving non-linear pattern recognition problems not only in the field of computer science, but also in high energy physics\cite{LSTMMagnetic,Atlasneural,BinaryFuyue}. Hence, we propose a novel algorithm that uses a fully-connected neural network to find out the patterns from the signal waveform and calculate the particle arriving time. 

The network is built and trained based on the TMVA package\cite{Tmva2007} in ROOT\cite{Root1997}. It has an input layer which contains several uniformly distributed points extracted from the waveform of signal. The output is the length of the leading edge $t_l$. By obtaining the peak time $t_p$ and $t_l$, the estimated particle arriving time is defined to be $t_a=t_p-t_l$. There are about 5 hidden layers in the network, and each of them has $\sim$15 nodes. The mean squared error of the predictions and labels is used as the loss function. $f(x)=tanh(x)$ is chosen to be the activation functions for all the hidden layers, and $f(x)=x$ for the output layer. Networks of different structures are trained and validated with the training($>$120,000) and validation($>$50,000) datasets. Both of these two datasets are from the simulation. All the networks used in this paper converge at last and no signs of overfitting are found, so normalization term is not added to the loss function.

\section{Result}
\label{sec:resu}
\subsection{\label{sec:ressimu}The result of test on simulation data}
To explore the optimal time resolution of the 0.104 $mm$ MRPC detector, the current signal in the simulation is readout by a fast front-end electronics, and length of the leading edge is around 1 $ns$.  The input of the network used to test the simulation data is the amplitude of 13 uniformly distributed points along the signal leading edge. The network gives an estimation of the $t_l$ and the time resolution is defined as:
\begin{equation}
\label{eq:NNfcn}
\sigma_{simu}=\sigma(t_{a,estimated}-t_{a,true})
\end{equation}

\begin{figure}
    \centering
    \begin{subfigure}[b]{0.45\textwidth}
        \includegraphics[width=\textwidth]{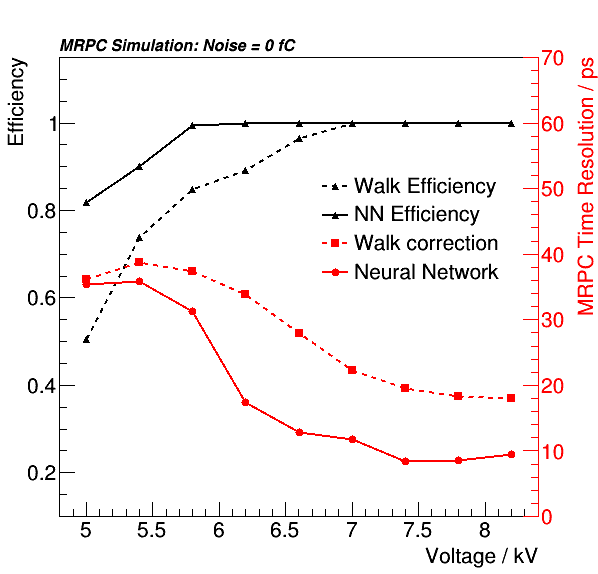}
        \caption{}
        \label{fig:voltage1}
    \end{subfigure}
    \begin{subfigure}[b]{0.45\textwidth}
        \includegraphics[width=\textwidth]{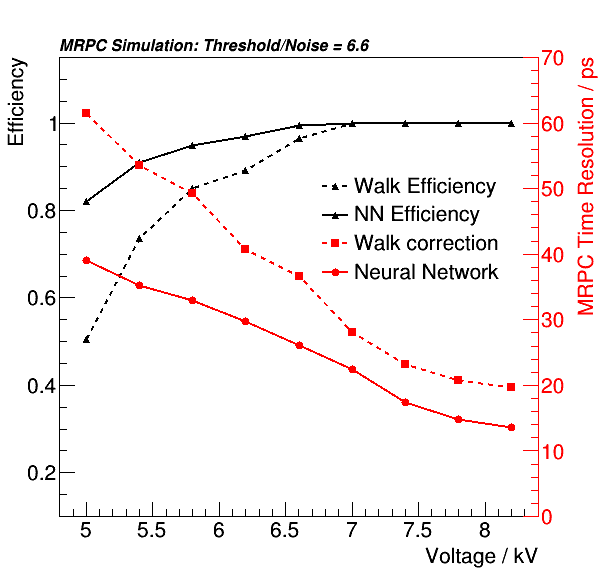}
        \caption{}
        \label{fig:voltage2}
    \end{subfigure}
    \caption{Efficiency and time resolution versus the high voltage added to each stack (a)The electronics noise is 0; (b)Threshold/Noise is 6.6.}
    \label{fig:voltage}
\end{figure}

Fig.\ref{fig:voltage} shows the efficiency and time resolution with respect to the high voltage added to each stack. The doted lines represent the result of the ToT method, while the solid lines are from NN. The uncertainty of the cosmic muons arriving time $\sigma(t_{a,true})$ is 30 $ps$ and is subtracted in the quadrature form. Fig.\ref{fig:voltage}(a) shows the performance of an optimal MRPC with electronics noise set to be 0, while the noise of Fig.\ref{fig:voltage}(b) is $1/6.6$ of the threshold. The efficiency nearly reaches 1 when the voltage is greater than 7 $kV$. NN has a higher efficiency compared to the ToT method, because it learns the pattern of all the read out waveforms including very small ones. However when the signal is tremendously large or small, these cases do not appear very often in the training data, and the resolution can be very large. Hence, these events are excluded and the efficiency becomes smaller than 100\%. 

The optimal time resolution for ToT and NN method is 20 $ps$ and 10 $ps$ respectively, and NN always performs better than the ToT method. When the voltage is lower than 6.5 $kV$, the noise largely degrades the time resolution. This is because in this region, the avalanche size inside the gas gap is small and thus induces a small signal that can be easily affected by the electronics noise. Under this condition, the NN algorithm, which utilizes the information of the entire leading edge, is more robust and achieves a much better precision. 
\subsection{\label{sec:ressimu}The result of test on experiment data}
\begin{figure*}[h!]
	\centering
	\includegraphics[width=0.45\textwidth]{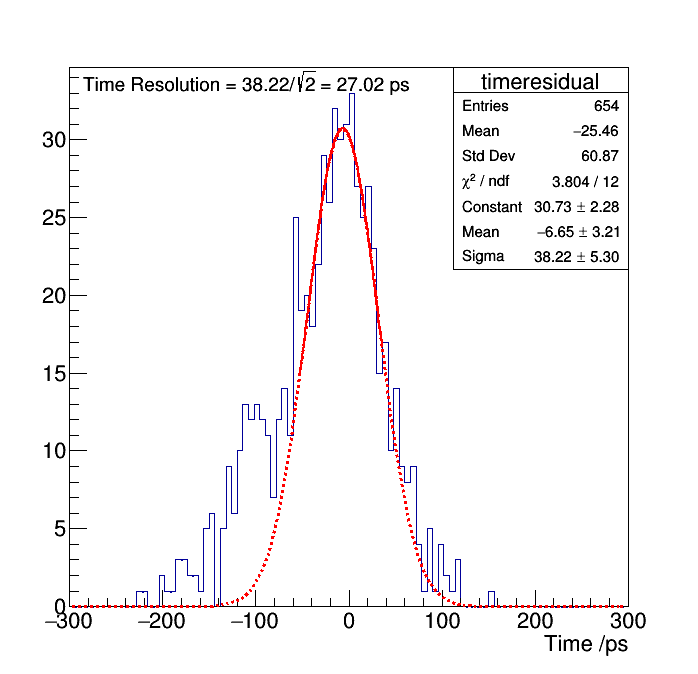}
	\caption{Time resolution of the test on experiment data}
	\label{fig:expe}
\end{figure*}

To test how well the neural network performs on the experiment data, we made 2 identical MRPCs which are exactly the same as the one used in the simulation. MRPCs are placed one on top of the other, and connect to a fast amplifier and a waveform digitizer respectively. Two PMTs which are placed on top and bottom of the 2 MRPCs are used as the trigger. The sampling rate is 5 $GHz$, and about 8 points that are uniformly distributed along the leading edge are obtained. MRPCs are read out from both sides, and therefore 4 waveforms(1L,1R,2L,2R) will be obtained for every single event. To be in agreement with the experiment, the simulated waveforms also contain 8 points along the leading edge. Taking the input of these 8 points, the neural network gives an estimation of $t_l$. For every event, $t_a$ is calculated separately for the 4 waveforms(1L,1R,2L,2R), and the estimated time for each MRPC($t_{est1}$,$t_{est2}$) is defined to be the average of the left and right. We calculate the difference of the two estimated MRPC time to eliminate the influence of the trigger:
\begin{equation}
\label{eq:deltat}
\Delta t=t_{est2}-t_{est1}
\end{equation}
Since the difference of 2 MRPC's truth time $t_{true1,2}$ is a constant, the uncertainty of $\Delta t$ is:
\begin{equation}
\label{eq:deltaresidual}
\sigma(\Delta t)=\sigma(t_{est2}-t_{true2}+t_{true1}-t_{est1})=\sigma(t_{res1}-t_{res2})
\end{equation}
\noindent where $t_{res1,2}$ are the time residual for 2 MRPCs and they are independent of each other. Then:
\begin{equation}
\label{eq:mrpcreso}
\sigma(\Delta t)=\sigma(t_{res1}-t_{res2})=\sqrt{\sigma^2(t_{resi1})+\sigma^2(t_{resi2})}=\sqrt{2\sigma^2_{MRPC}}
\end{equation}
\begin{equation}
\label{eq:mrpcreso}
\sigma_{MRPC}=\frac{\sigma(\Delta t)}{\sqrt{2}}
\end{equation}
\indent Fig.\ref{fig:expe} shows the distribution of $\Delta t$. The time resolution of MRPC obtained from the neural network is 27 $ps$. To compare with the ToT method, we set a threshold to the experiment waveforms and discriminate $t_c$ and $t_{tot}$. After the slewing correction, the best result we achieve is only 48 $ps$. This effectively proves that the neural network is a promising analysis algorithm to reconstruct the MRPC time. 
\section{Conclusions}
\label{sec:concl}
We have proposed a novel idea to introduce the neural network algorithm to reconstruct the time of the MRPC detector. For the experiment data, the best result obtained from the neural network is around 27 $ps$ for a 104$\mu m$-thick MRPC, which is a lot better than the result of the ToT method. NN also shows its great power at low voltage when the electronics noise exists. Further research on the pattern of the signal and the structure of the neural network is being processed. Therefore the optimal time resolution is expected to be even better.



\bibliographystyle{elsarticle-num}
\bibliography{myrefs.bib}{}

\end{document}